\documentclass[11pt,twoside]{article}
\usepackage{asp2014}

\aspSuppressVolSlug
\resetcounters
\begin{document}

\title{Nonlinear fitting of undersampled discrete datasets in astronomy}

\author{Igor~Chilingarian,$^1$ and Kirill~Grishin$^2$}
\affil{$^1$Center for Astrophysics --- Harvard and Smithsonian, 60 Garden Street MS09, Cambridge, MA 02138, USA}
\affil{$^2$Universit\'e Paris Cit\'e, CNRS(/IN2P3), Astroparticule et Cosmologie, 10 rue Alice Domon et L\'eonie Duquet, 75013, Paris, France}

\paperauthor{Igor~V.~Chilingarian}{igor.chilingarian@cfa.harvard.edu}{0000-0002-7924-3253}{Center for Astrophysics --- Harvard and Smithsonian}{}{Cambridge}{MA}{02138}{USA}
\paperauthor{Kirill~A.~Grishin}{kirill.grishin@voxastro.org}{0000-0003-3255-7340}{Universit\'e Paris Cit\'e, CNRS(/IN2P3)}{Astroparticule et Cosmologie}{Paris}{}{75013}{France}

\begin{abstract}
Data analysis and interpretation often relies on an approximation of an empirical dataset by some analytic functions or models. Actual implementations usually rely on a non-linear multi-dimensional optimization algorithm, typically Levenberg--Marquardt (LM) or other flavors of Newtonian gradient methods. A vast majority of datasets in optical and infrared astronomy are represented by values on a discrete grid because the actual signal is sampled by regularly shaped pixels in the light detectors. Here we come to the main problem of nearly all widely used implementations of nonlinear optimization methods: the function that is being fitted is evaluated at central pixel positions rather than integrated over the pixel areas. Therefore, the best-fitting set of parameters returned by the minimization routine might not be the best representation of the observed dataset, especially if a dataset is undersampled. For example, a central pixel of a 1D Gaussian with a dispersion of 1~pix  (2.36~pix FWHM; so not too strongly undersampled) will be about 4.2\% lower than its central evaluated value if integrated. To handle this effect properly, one needs to perform numerical or analytic integration of a model within the pixel boundaries. We will discuss possible computationally efficient solutions and test our preliminary implementation of a nonlinear fitting using LM minimization that correctly accounts for the discrete nature of the data. 
\end{abstract}

\section{Motivation}
Many scientific applications in astronomy rely on accurate measurements of specific properties of the signal from the data. For example, computing fluxes and widths of emission lines in spectra or fluxes and sizes of stars and galaxies in images. It is also well known that if one uses a known shape of the instrumental response (a line spread function (LSF) in the case of spectra or a point spread function (PSF) in the case of images) to measure fluxes of unresolved lines/sources, the final measurement quality significantly improves (\citealp{1986PASP...98..609H}; a.k.a. optimal extraction for more than a century). These operations require an approximation of an observed signal by a model. If the fitting is done in the pixel space (Fourier-space fitting is also possible in some cases but requires special efforts for accurate treatment of errors), the evaluation of the model function is performed on a pixel grid. Sometimes the PSF/LSF or the model function itself is undersampled, i.e. does not satisfy the Nyquist--Shannon criterion.

The main problem when fitting undersampled data arises from the fact that most (if not all) widely used implementation of both linear and nonlinear fitting algorithms evaluate the fitting function in the pixel centers, while instead it should be integrated within the pixel borders (or the pixel area in the two-dimensional case).

\begin{figure}
\includegraphics[width=\hsize]{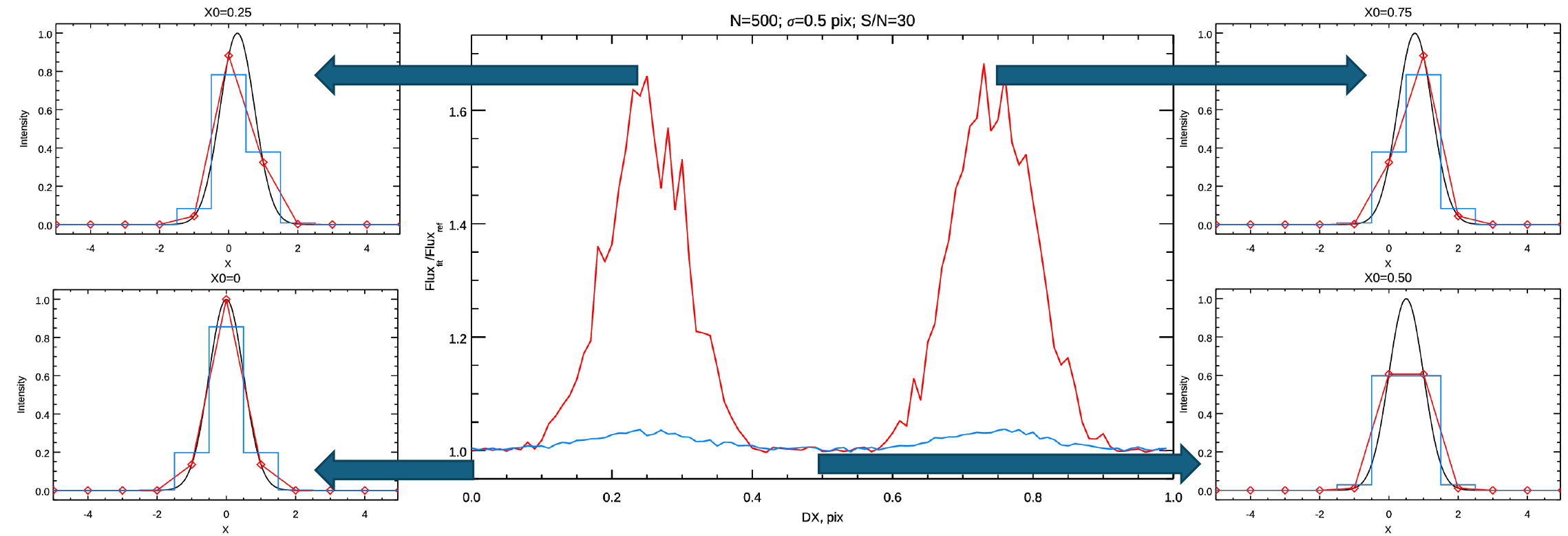}
\caption{{\bf Side Panels:} Undersampled 1D Gaussian functions with $\sigma$=0.5~pix (FWHM=1.18~pix) shifted by DX=0 (bottom left), DX=0.25 (top left), DX=0.50 (bottom right), DX=0.75 (top right). Black lines show “the true shapes”, red lines show functions sampled at pixel centers, blue step-line display integrated values within pixels. {\bf Central panel:} a Monte-Carlo simulation (500 noise realizations that correspond to S/N=30 in the brightest pixel) of the recovery of the flux by a nonlinear Levenberg-Marquardt minimization for the standard evaluation of a Gaussian function in pixel values (red) and integrated over pixel areas (blue). The true flux value is 1. \label{P909_fig1}}
\end{figure}

\begin{figure}
\includegraphics[width=0.49\hsize]{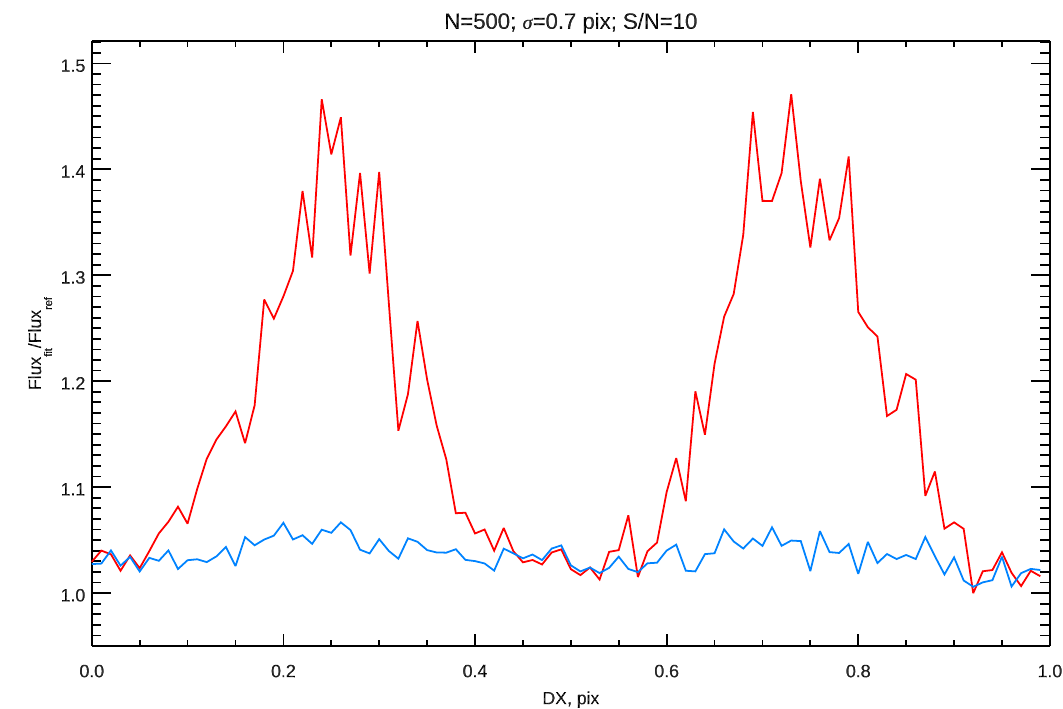}
\includegraphics[width=0.49\hsize]{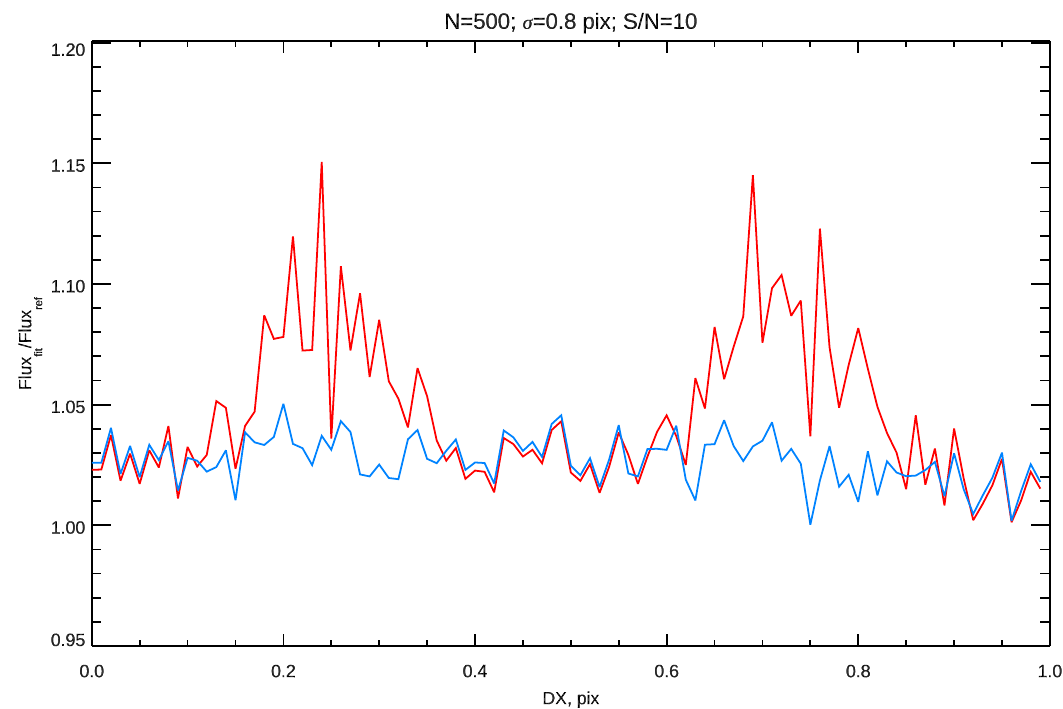}
\includegraphics[width=0.49\hsize]{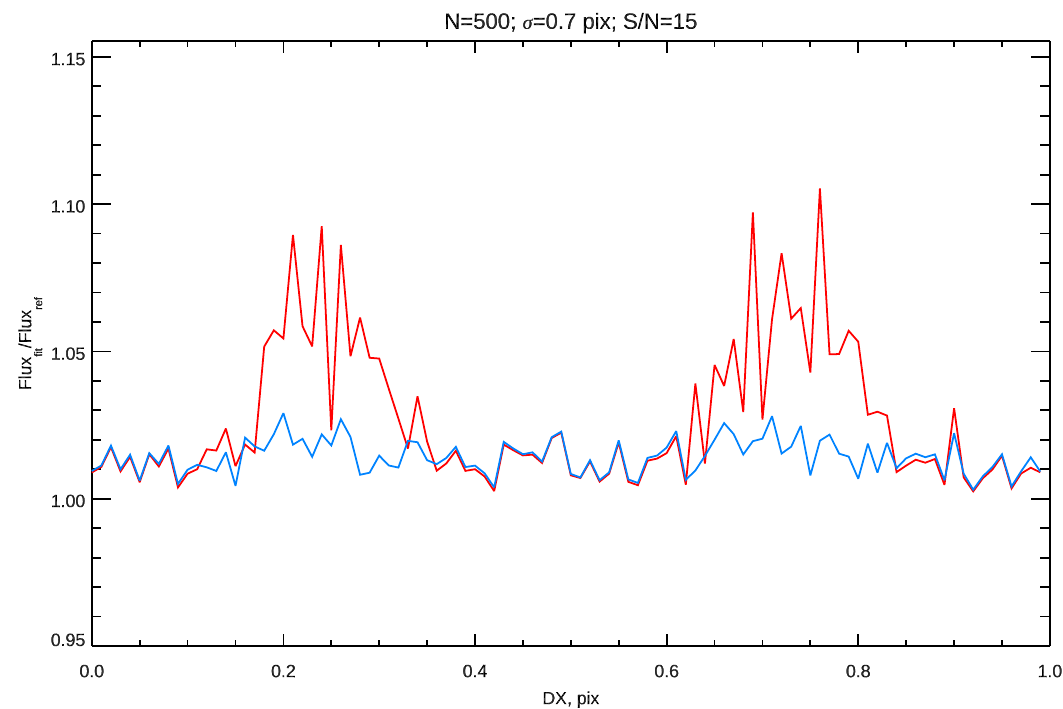}
\includegraphics[width=0.49\hsize]{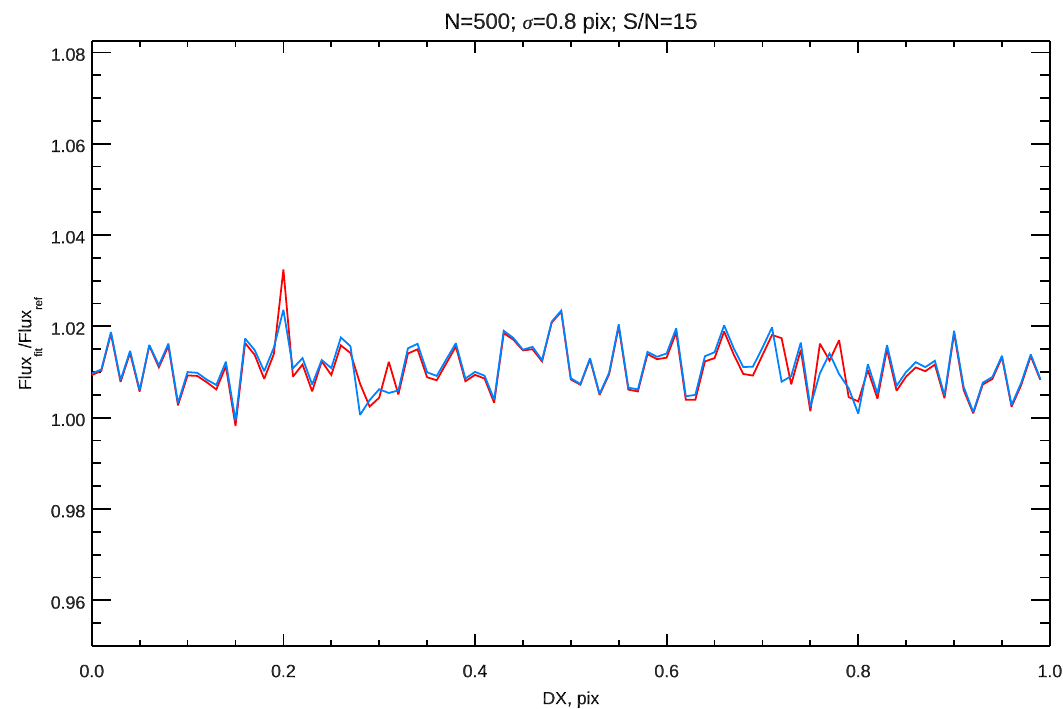}
\caption{The signa-to-noise ratio effect (top: 10, bottom: 15) on the quality of flux recovery at 87\%\ (left) and 100\%\ (right) of the Nyquist sampling. Red lines: sampling at pixel centers; blue lines: integration within pixels.
\label{P909_fig2}}
\end{figure}

\section{Potential Solutions}

One solution is to increase the signal-to-noise ratio of the data (or, effectively, the exposure time). However, SNR typically scales as t$^{1/2}$, therefore, providing a typical oversubscription rate of observing time of 7--10 at the most demanded facilities (Hubble Space Telescope; James Webb Space Telescope), increasing the SNR might not be a viable solution.

Alternatively, we can modify the fitting algorithm so that it operates on data integrated within one- or multi-dimensional pixels. In special cases, when the integrals of the function that represents a model can be computed analytically, this becomes straightforward to implement. Moreover, the computation of the Jacobian matrix that substantially increases the efficiency and performance of a typical Newtonian-like gradient method also becomes straightforward to calculate. For example, in the case of a Gaussian, the integral is expressed as an error function that is available in all modern programming languages. The improvement in the flux recovery for low SNR data in undersampled cases is drastic (see Fig.~\ref{P909_fig1}--\ref{P909_fig2} presenting our implementation; blue lines). Even a 13~\%-undersampled (w.r.t Nyquist) Gaussian emission line at SNR$=$10 can lead to an overestimate of the line flux by a factor of 1.5, while replacing the evaluation with integration within the pixel boundaries reduces the bias factor to 1.02.

If analytic integration is impossible, then numerical integration methods should be used instead, which, however, become computationally expensive when the dimensionality of the data increases. In addition, it is suggested to evaluate the model function on an oversampled grid to reduce numerical integration artefacts.

\section{Implications and Notes on Implementation}

\subsection{1D: spectral emission lines}

Some single-order spectrographs by design have an undersampled LSF (usually in the blue end of the wavelength coverage). Perhaps, the most well-known example is the MUSE integral field unit (IFU) spectrograph \citep{2010SPIE.7735E..08B} operated at ESO VLT, which has a constant sampling of 1.25~\AA/pix and the FWHM spectral resolution of about 2.4~\AA. Effectively, if one tries to measure individual emission lines, the signal-to-noise floor becomes 7--10 per pix. 

On the other hand, when parameters are extracted from an absorption-line spectrum, moderate undersampling does not lead to serious numerical artefacts because the spectral line shape is sampled by numerous absorption features. \citet{2020PASP..132f4503C} demonstrated that the key factor that defines the accuracy of the extracted kinematic parameters are the depths and contrast of absorption lines.

\subsection{2D: point sources (e.g. stars)}

Although 2D is less affected by undersampling compared to 1D, the situation becomes crucial for space missions, where the PSF is sometimes strongly undersampled. In fact, all three major space observatories operating in the optical/near-infrared domains and delivering imaging data, HST, JWST, and Euclid all have undersampled PSFs. For example, JWST NIRCam \citep{2023PASP..135b8001R} has FWHM$=0.8$~pix in the \textit{F090W} filter widely used in resolved stellar population studies. Similar PSF undersampling factors are not uncommon in IFU spectrographs operated at large telescopes in both seeing-limited \citep[e.g. Binospec-IFU at the 6.5~m MMT;][]{2025arXiv250101528F} and adaptive optics-assisted \citep[e.g. ERIS at ESO VLT;][]{2023A&A...674A.207D} modes.

In some cases, direct numerical 2D integration may become more expensive in terms of computational resources, compared to a simple evaluation of a function in centers of pixels. Moreover, in the case of nonstandard pixel shape, such as IFU spectrographs with hexagonal lenslets (e.g. Binospec-IFU) or round fibers (e.g. PMAS/PPAK), even the functions which can be integrated analytically (e.g. Gaussian) become difficult to integrate. In such cases, the use of relations between the area and edge integrals can substantially simplify the calculation and reduce the dimensionality of the problem. For example, in the case of 2D datasets, the use of Green's theorem leads to the transformation of a 2D integral over the area to 1D integration along the edge of a pixel. According to Green's theorem:
\begin{equation}
\iint_R \left( \frac{\partial Q}{\partial x} - \frac{\partial P}{\partial y} \right) \, dxdy = \int_C \left( P \, dx + Q \, dy \right)
\end{equation}
To perform this transformation, one should find $P(x, y)$ and $Q(x,y)$, which should satisfy $\frac{\partial Q}{\partial x} - \frac{\partial P}{\partial y} = f(x,y)$, where $f(x,y)$ is a flux distribution. These functions can be found by solving first-order differential equations. For a 2D Gaussian, they are:
$$
P(x, y) = -\frac{1}{4 \sigma \sqrt{2 \pi}} \exp\left(-\frac{1}{2} \left( \frac{x - x_c}{\sigma} \right)^2 \right) 
\cdot \text{erf}\left(\frac{y - y_c}{\sigma\sqrt{2}} \right)
$$
$$
Q(x, y) = \frac{1}{4 \sigma \sqrt{2 \pi}} \exp\left(-\frac{1}{2} \left( \frac{y - y_c}{\sigma} \right)^2 \right) 
\cdot \text{erf}\left(\frac{x - x_c}{\sigma\sqrt{2}} \right)
$$

The use of this approach on large pixel grids is easy to parallelize and can save computation time by up-to 50~\%\ because $n$ tightly packed hexagonal pixels will have $3n$ unique edges (versus $6n$ total), which require calculation of integrals (the same applies to rectangular pixels that have $2n$ unique edges).

\acknowledgments

IC’s research is supported by the SAO Telescope Data Center. IC acknowledges support from the NASA ADAP-22-0102 grant. KG acknowledges support from ANR-24-CE31-2896.

\bibliography{P909}


\end{document}